\documentclass[11pt]{article}

\usepackage[margin=1in]{geometry}
\usepackage{cite}
\usepackage{amsmath,amssymb,amsfonts}
\usepackage{graphicx}
\usepackage{booktabs}
\usepackage{url}
\usepackage[hidelinks]{hyperref}
\usepackage{microtype}

\graphicspath{{figures/}}

\title{EA-Ops: Git-Native Architecture as Code for Continuous Enterprise Architecture Governance}
\author{%
Vahid Tavakkoli$^{1}$, Kabeh Mohsenzadegan$^{1}$, and Kyandoghere Kyamakya$^{1,2}$\\[0.6em]
\small $^{1}$Department of Smart Systems Technologies, University of Klagenfurt, 9020 Klagenfurt am W\"orthersee, Austria\\
\small (e-mail: vahid.tavakkoli@aau.at; kabeh.mohsenzadegan@aau.at; \\
\small kyandoghere.kyamakya@aau.at)\\
\small $^{2}$Facult\'e Polytechnique, Universit\'e de Kinshasa, Kinshasa, Democratic Republic of the Congo\\[0.4em]
\small Corresponding author: Vahid Tavakkoli (e-mail: vahid.tavakkoli@aau.at).
}
\date{}

\begin{document}
\maketitle

\begin{abstract}
Enterprise architecture (EA) repositories frequently separate architecture models from the engineering workflow used to change software and infrastructure. This article presents EA-Ops, an open-source Git-native Enterprise Architecture-as-Code framework that represents architecture facts as YAML, validates typed relationships against an ArchiMate 3.2 profile, enforces organization-specific governance rules, performs graph-based change-impact analysis, and publishes human-facing reports and a static interactive portal from the same reviewed source. We evaluate EA-Ops with a reproducible GitHub Actions harness. Eight independently injected structural, semantic, and governance fault classes were executed across 30 trials each; all 240 trials matched ground truth exactly, with precision, recall, and $F_1$ of 1.000. Scalability experiments with 30 measured repetitions reached 50,000 objects and 100,000 relationships: median validation time was 52.582~s, median impact traversal was 627.000~ms, and peak resident-set size was 919.2~MB. A ten-scenario Metroville digital-permit reference architecture produced exact validation outcomes and exact impact-set agreement with an independent breadth-first-search oracle in every scenario. The configured 100,000-object end-to-end benchmark generated its model successfully but exceeded the 180-minute CI budget during the performance stage; no timing result is extrapolated. At 50,000 objects, Markdown report generation rather than semantic validation is the dominant scaling bottleneck. The results support Git-native continuous governance as a practical EA operating model at tens-of-thousands-of-object scale while defining clear limits and optimization targets for larger repositories.
\end{abstract}


\section{Introduction}
\label{sec:introduction}
Enterprise architecture (EA) is expected to maintain a coherent description of an organization's business, information, application, and technology landscape and to support decisions about change. ISO/IEC/IEEE 42010:2022 formalizes concepts for architecture descriptions, while ArchiMate 3.2 and the TOGAF Standard provide widely used foundations for enterprise-level modeling and architecture practice \cite{iso42010,archimate32,togaf10}. The challenge is that an architecture description is valuable only while it remains synchronized with the systems, dependencies, and governance decisions it is intended to represent. Recent EA studies continue to examine modeling quality, value representation, reference architectures, technical dependency discovery, and the evolution toward digital representations of organizations \cite{sanyoto2023,egeten2023,edrisi2024,goyal2024,shankaravelu2024,utami2024,nadobny2024}.

Software delivery has meanwhile adopted version-controlled and continuously validated operating models. GitOps treats Git as a desired-state source and couples reviewed changes with automated reconciliation \cite{beetz2022,gupta2022,lopez2022,kurrewar2025}. Infrastructure as Code (IaC) moves operational definitions into machine-readable artifacts, enabling automated testing and static analysis but also exposing new correctness and maintenance concerns \cite{golis2022,begoug2023,sokolowski2024,bessghaier2024}. CI/CD research further emphasizes secure, observable, and continuously checked delivery workflows \cite{bajpai2022,borges2025}.

These developments motivate a corresponding operating model for architecture. Recent software-architecture research explicitly studies \emph{Architecture as Code} (AaC) as version-controlled, machine-readable architecture descriptions \cite{bucaioni2025,pontillo2026}. Architecture knowledge is also increasingly captured and analyzed through lightweight decision records and semantic representations \cite{buchgeher2023,karetnikov2024}, while architecture reconstruction, change-impact analysis, model-driven engineering, and model-driven DevOps provide complementary mechanisms for keeping architecture and implementation aligned \cite{cerny2022,cerny2025,marcen2024,karlovs2025,khadem2025,sawant2026}.

A direct transfer of GitOps or IaC practices to EA is nevertheless incomplete. Enterprise architecture repositories contain typed cross-layer relationships, organization-specific policy, multiple stakeholder views, and dependency questions that generic configuration validation does not capture. An EA-as-code system must preserve architectural semantics, make governance executable, expose change consequences before merge, and produce stakeholder-facing views without creating a second source of truth. It should also provide empirical evidence that its checks are deterministic and that its workflow remains operational at repository scales beyond toy examples.

This article presents \emph{EA-Ops}, an open-source Git-native Enterprise Architecture Operations framework. EA-Ops stores architecture facts as YAML; validates identifiers, references, element types, and ArchiMate-aware relationship combinations; evaluates organization-defined governance rules; computes graph-based change impact; and generates reports and a static interactive portal from the same reviewed source. Pull requests act as architecture change requests, while the main branch represents the reviewed architecture state. The implementation is deliberately lightweight: Python and PyYAML implement the model-processing pipeline, and the portal is emitted as static HTML, JavaScript, and SVG.

The article addresses four research questions:
\begin{itemize}
	\item \textbf{RQ1 -- Validation correctness:} How accurately does EA-Ops detect controlled structural, semantic, and governance faults against independently declared ground truth?
	\item \textbf{RQ2 -- Scalability:} How do validation time, graph-impact latency, peak memory, and publication time change as deterministic synthetic EA models grow from 100 to 50,000 fully measured objects?
	\item \textbf{RQ3 -- Controlled change analysis:} Does EA-Ops produce the expected validation outcome and the same impact set as an independent graph oracle for realistic cross-layer architecture changes?
	\item \textbf{RQ4 -- Reproducibility:} Can the evaluation itself be executed as a versioned CI workflow that records environment metadata and publishes raw artifacts suitable for independent inspection?
\end{itemize}

The contributions are fourfold. First, we define and implement an integrated Git-native EA operating model combining semantic EA modeling, governance-as-code, pull-request review, impact analysis, and publication. Second, we provide an independent fault-injection evaluation covering eight fault classes and 240 trials. Third, we characterize scalability with repeated measurements up to 50,000 objects and 100,000 relationships and report the unsuccessful 100,000-object end-to-end run rather than extrapolating an unobserved result. Fourth, we evaluate ten controlled changes in the Metroville reference architecture against both explicit validation expectations and an independent breadth-first-search impact oracle. The software, benchmark harness, case-study repository, raw measurements, and workflow metadata are public and reproducible.

The remainder of this article reviews related work, defines the EA-Ops model and workflow, describes the evaluation, reports the results, discusses implications and threats to validity, and concludes with reproducibility guidance and future work.

\section{Background and Related Work}
\label{sec:related}
\subsection{Enterprise Architecture Standards and Executable Models}
Architecture-description practice is anchored by ISO/IEC/IEEE 42010:2022, which distinguishes architectures from their descriptions and defines stakeholders, concerns, viewpoints, and model kinds \cite{iso42010}. For enterprise-level modeling, ArchiMate 3.2 supplies a standardized language spanning strategy, business, application, technology, motivation, and implementation domains, while TOGAF provides a broader architecture method and governance framework \cite{archimate32,togaf10}. Recent studies investigate ArchiMate's strengths and limitations as an EA modeling language \cite{sanyoto2023}, the role of enterprise-architecture frameworks in governance contexts \cite{egeten2023}, value modeling in ArchiMate \cite{shankaravelu2024}, and reference architectures for digital-government settings \cite{utami2024}. Other work demonstrates the usefulness of machine-readable EA models for discovering technical dependencies during legacy transformation \cite{nadobny2024}, explores natural-language assistance for EA analysis \cite{goyal2024}, and maps EA concepts into evolvable digital twins of organizations \cite{edrisi2024}. These studies reinforce the value of explicit architecture models but do not by themselves define a Git-native continuous governance lifecycle.

\subsection{GitOps, Infrastructure as Code, and Continuous Validation}
GitOps extends DevOps by making declarative desired state and Git-centered change control central to operational workflows \cite{beetz2022}. Empirical and applied studies report growing GitOps use in CI/CD \cite{gupta2022}, demonstrate GitOps deployment patterns in edge computing \cite{lopez2022}, and apply GitOps to Kubernetes deployment automation \cite{kurrewar2025}. IaC research addresses a closely related problem at the infrastructure layer. Golis \emph{et al.} combine microservices, CI/CD, and IaC for dynamic container creation \cite{golis2022}; Begoug \emph{et al.} analyze the practical issues IaC users discuss at scale \cite{begoug2023}; Sokolowski \emph{et al.} propose automated testing for IaC programs \cite{sokolowski2024}; and Bessghaier \emph{et al.} study the prevalence and impact of IaC smells \cite{bessghaier2024}. CI/CD security and observability studies similarly show that automated pipelines require explicit controls and continuously evaluated quality attributes \cite{bajpai2022,borges2025}. EA-Ops adopts these engineering properties--versioned desired state, deterministic checks, and CI evidence--but applies them to architecture semantics and enterprise governance rather than deployment infrastructure.

\subsection{Architecture Knowledge, Architecture as Code, and Change Impact}
Versioned architecture knowledge is increasingly treated as an engineering artifact. Architecture Decision Records provide lightweight, repository-friendly records of architectural rationale, and empirical work has examined their use in open-source projects \cite{buchgeher2023}. Semantic modeling of ADRs further shows how architecture knowledge can be made machine-processable for automated analysis \cite{karetnikov2024}. At the structural level, architecture reconstruction research surveys static and dynamic techniques for recovering microservice architectures \cite{cerny2022}, and recent work develops infrastructure for cross-service change-impact analysis during system evolution \cite{cerny2025}. Lightweight executable architectural validation is also being explored through automated test generation \cite{sawant2026}.

Most directly, Bucaioni \emph{et al.} formalize and investigate \emph{Architecture as Code} as a software-architecture approach in which architecture artifacts are treated as code-like, version-controlled assets \cite{bucaioni2025}. Pontillo \emph{et al.} subsequently report a multiple-case empirical study of AaC in industrial settings \cite{pontillo2026}. EA-Ops does not claim that the general AaC concept is new. Its contribution is an open implementation and empirical evaluation of an \emph{enterprise}-architecture lifecycle that combines typed EA semantics, organization policy, pull-request governance, dependency impact, and publication from one repository.

\subsection{Model-Driven DevOps and the Remaining Gap}
Model-driven engineering (MDE) provides a long-standing basis for systematic use of models in software engineering; a recent review characterizes the increasing intersection between MDE and machine learning \cite{marcen2024}. Model-driven DevOps work goes a step further by transforming high-level models into CI/CD automation \cite{karlovs2025}, while recent IEEE Access research maps DevOps and microservice-architecture concepts into decision-support structures \cite{khadem2025}. These approaches illustrate the value of connecting architectural intent with operational automation.

The literature therefore supplies complementary pieces: standards define architecture-description semantics; EA studies demonstrate the value of structured enterprise models; GitOps and IaC establish reviewed, executable operating practices; ADR and reconstruction research make architecture knowledge increasingly machine-processable; and AaC establishes the broader principle of codifying architecture. The gap addressed in this work is a reproducible, continuously governed EA workflow that integrates those properties in one Git-native artifact and evaluates both correctness and scale end to end.

\section{EA-Ops Design}
\label{sec:design}
\subsection{Git-Native Repository Model}
An EA-Ops repository contains four principal artifact categories:
architecture objects, relationships, views, and rules. The root configuration
maps these categories to repository paths and selects a metamodel profile.
Each object is identified by a stable ID and contains a type, name, optional
description, and properties. Relationships are first-class typed edges with
source and target IDs. Views are derived or curated projections and may persist
layout positions. Rules express organizational constraints independently of
the underlying model.

We model an architecture repository as
\begin{equation}
	\mathcal{A} = (V,Q,E,\tau_V,\tau_E,P,R),
\end{equation}
where $V$ is the set of architecture objects, $Q$ is the set of relationship
connectors, and
\[
E \subseteq (V \cup Q) \times (V \cup Q)
\]
is the set of relationships. The mapping $\tau_V$ assigns architecture
objects to element types, $\tau_E$ assigns relationships to relationship
types, $P$ maps architecture objects to their properties, and $R$ is the
governance-rule set. The connector set $Q$ may be empty for metamodel profiles
that do not support relationship connectors.

The approved repository state is represented by the main branch, following the versioned desired-state principle used in GitOps workflows \cite{beetz2022}. Proposed architecture modifications are ordinary Git changes, which means Git provides identity, repository permissions, diffs, review history, branch protection, and merge provenance. EA-Ops adds architecture-specific processing rather than creating a parallel collaboration subsystem.

\begin{figure*}[t]
	\centering
	\includegraphics[width=0.93\textwidth]{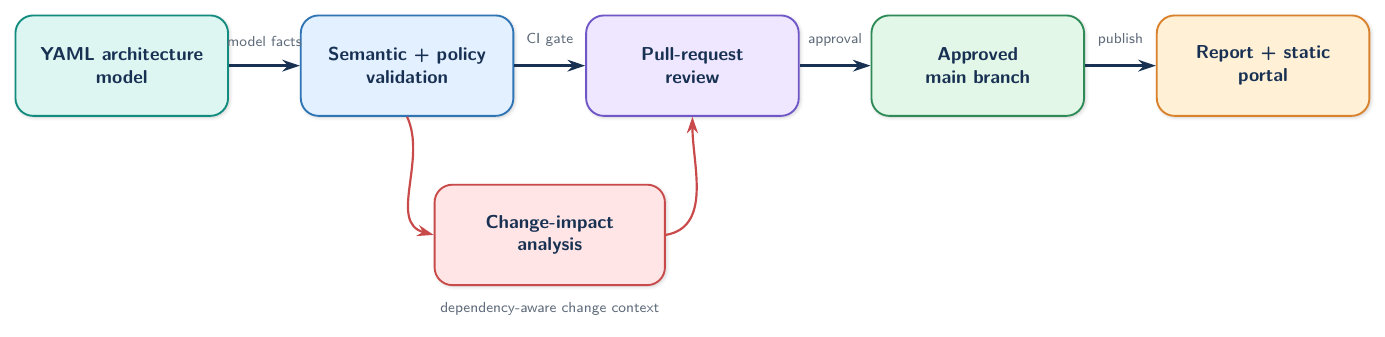}
	\caption{EA-Ops operating loop. Git supplies versioning and review mechanics; EA-Ops supplies architecture semantics, governance, impact analysis, and publication.}
	\label{fig:loop}
\end{figure*}

\subsection{Deterministic Validation}
Validation produces a set of findings
\begin{equation}
	F(\mathcal{A}) = F_{\mathrm{struct}} \cup F_{\mathrm{semantic}} \cup F_{\mathrm{gov}}.
\end{equation}
Structural findings include missing required fields, duplicate IDs, and dangling relationship endpoints. Semantic findings check element and relationship vocabulary and, when configured, validate each relationship against a pinned ArchiMate 3.2 relationship matrix. Governance findings are generated from repository rules. Current rule primitives include required properties, conditional property predicates, required relationship cardinalities, and allowed property-value sets.

This rule design separates organization policy from code. For example, an enterprise can require every application component to carry an owner and lifecycle, or restrict lifecycle to a controlled vocabulary, without changing the validator implementation. Error-severity findings can fail the architecture gate in CI; warnings can expose governance debt without blocking the change.

\subsection{Graph-Based Change Impact}
For a set of changed IDs $C$, the current EA-Ops impact operation traverses relationships in both directions and returns the connected set reachable from $C$ in the undirected projection of the architecture graph:
\begin{equation}
	I(\mathcal{A},C)=\{v \mid \exists c\in C: c \leadsto v\}.
\end{equation}
The implementation repeatedly scans all relationships for each frontier wave. If $D$ is the number of traversal waves and $|E|$ the number of relationships, the current implementation is bounded by $O(D|E|)$ for the traversal portion. This is intentionally reported because the evaluation concerns the implemented system, not an idealized adjacency-list implementation. Typed, directional, and depth-bounded impact semantics are natural future extensions.

\subsection{Publication and Human-Facing Views}
The same model generates a Markdown architecture report and a static interactive portal, retaining the machine-readable-model emphasis found in recent EA dependency and digital-twin work \cite{nadobny2024,edrisi2024}. The portal supports catalog search, cross-layer navigation, process views, governance findings, and draggable layout drafts that can be exported back to YAML. No database or server-side web application is required to browse the generated site. This preserves a single reviewed source of truth: browser layout changes remain local drafts until committed through Git.

\begin{figure*}[t]
    \centering
    \begin{minipage}[t]{0.49\textwidth}
        \centering
        \includegraphics[width=\linewidth]{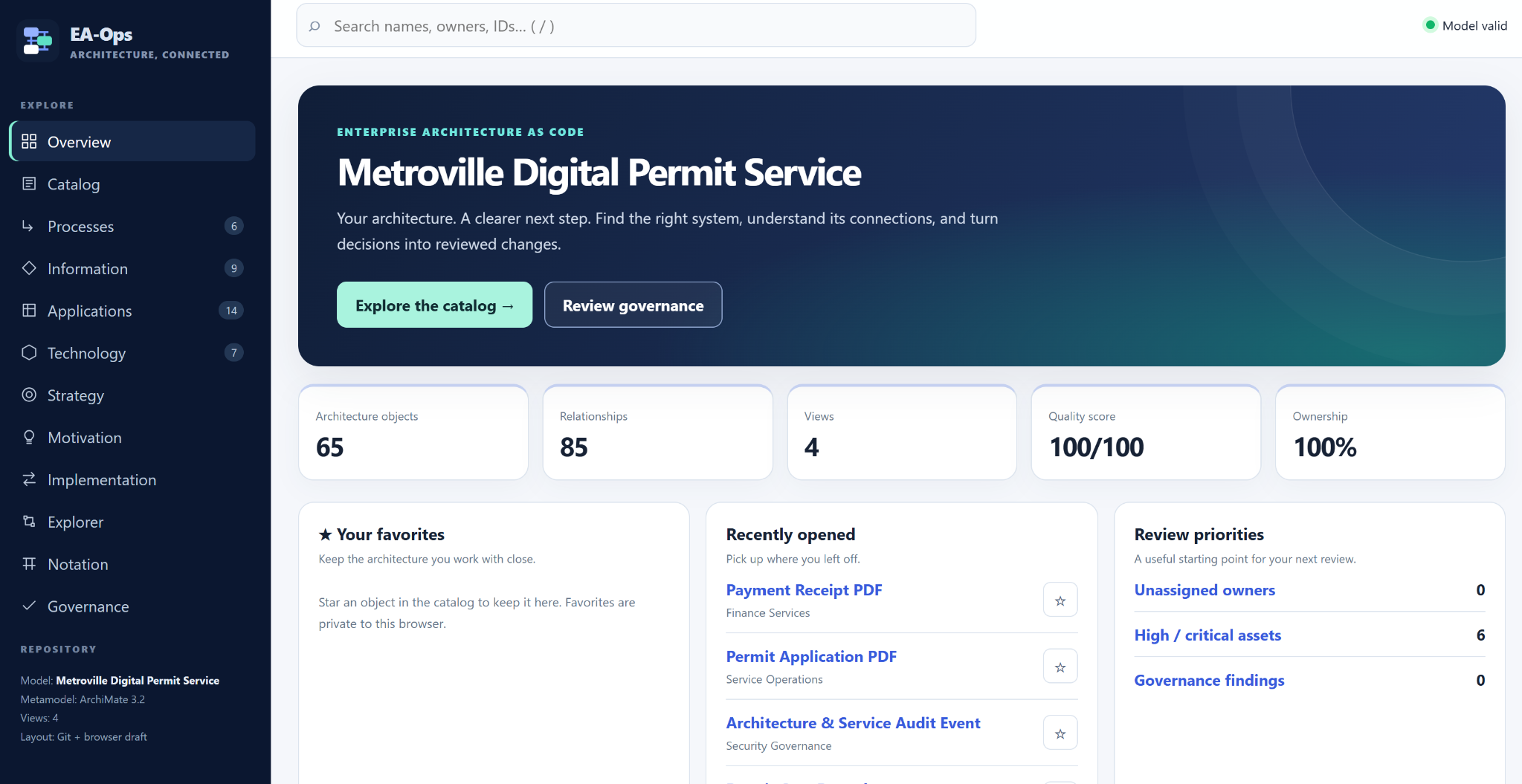}\par\vspace{0.6mm}
        {\footnotesize (a) Repository overview and architecture-health summary.}
    \end{minipage}
    \hfill
    \begin{minipage}[t]{0.49\textwidth}
        \centering
        \includegraphics[width=\linewidth]{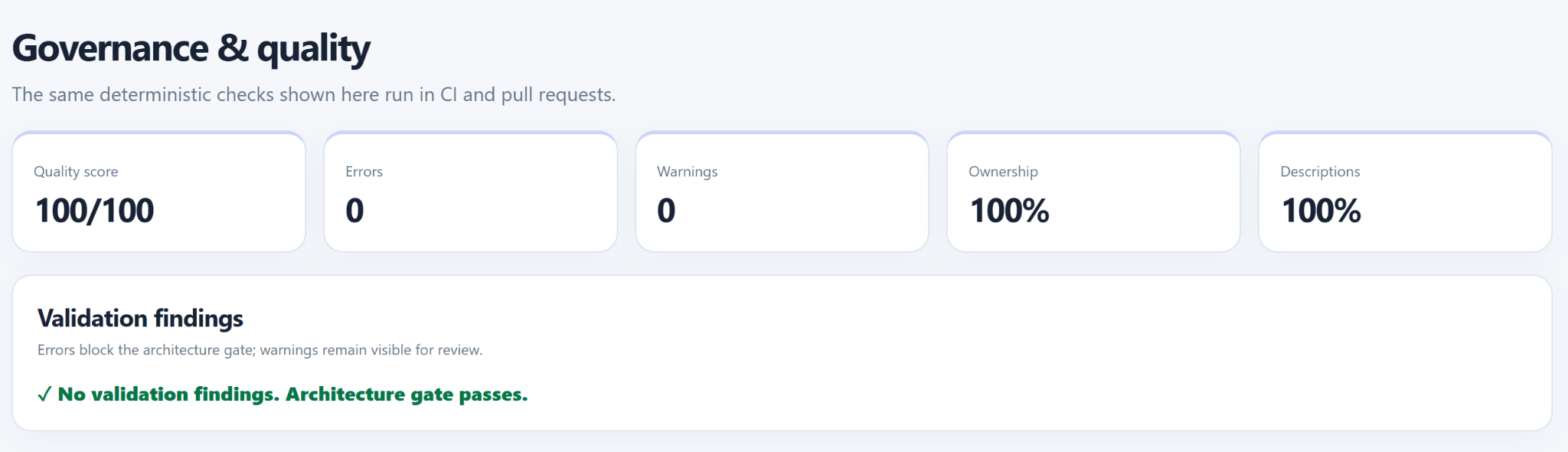}\par\vspace{0.6mm}
        {\footnotesize (b) Governance and quality gate using CI-equivalent checks.}
    \end{minipage}

    \vspace{2.2mm}

    \begin{minipage}[t]{0.49\textwidth}
        \centering
        \includegraphics[width=\linewidth]{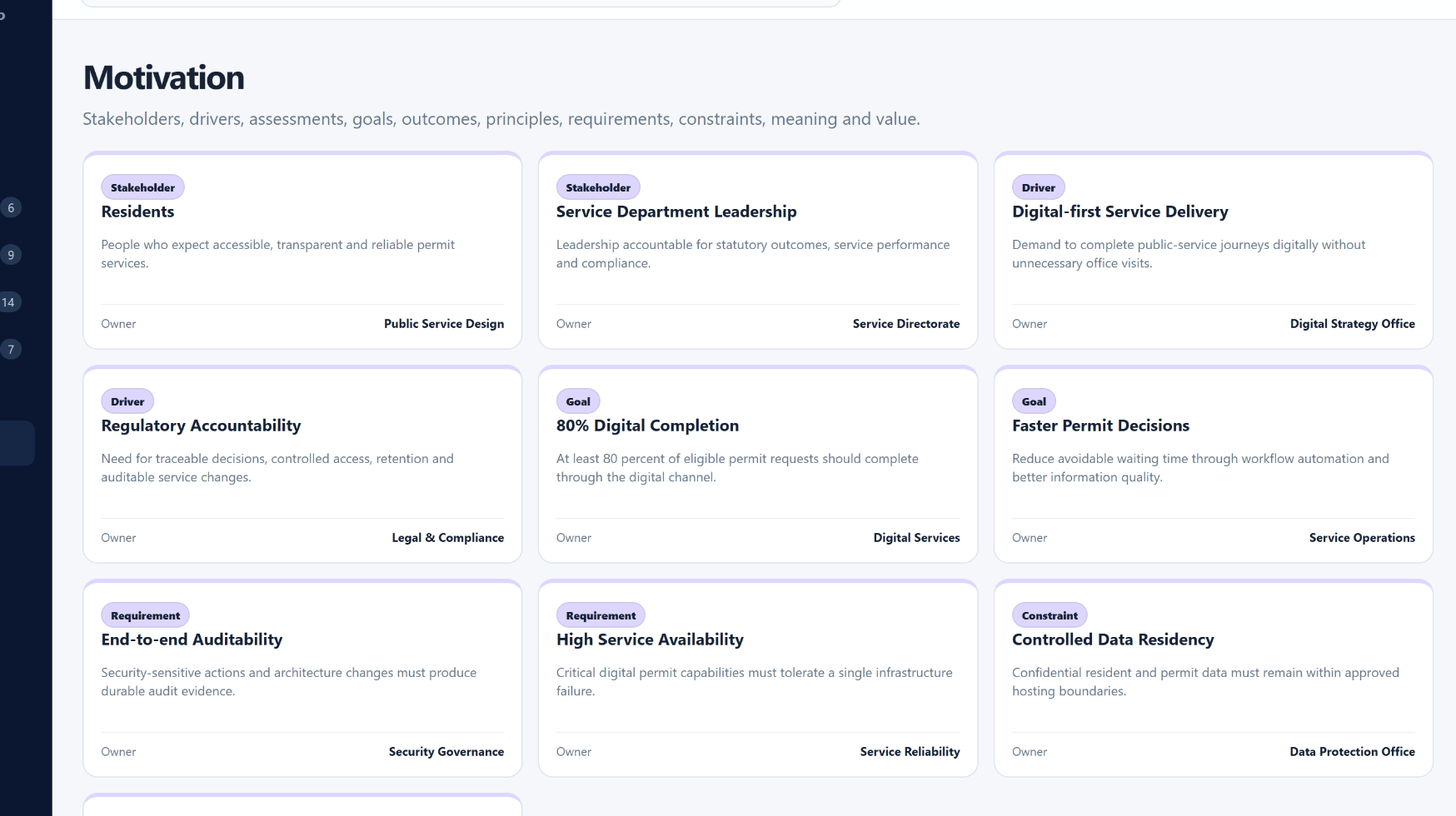}\par\vspace{0.6mm}
        {\footnotesize (c) Motivation catalog with stakeholders, drivers, goals, requirements, and constraints.}
    \end{minipage}
    \hfill
    \begin{minipage}[t]{0.49\textwidth}
        \centering
        \includegraphics[width=\linewidth]{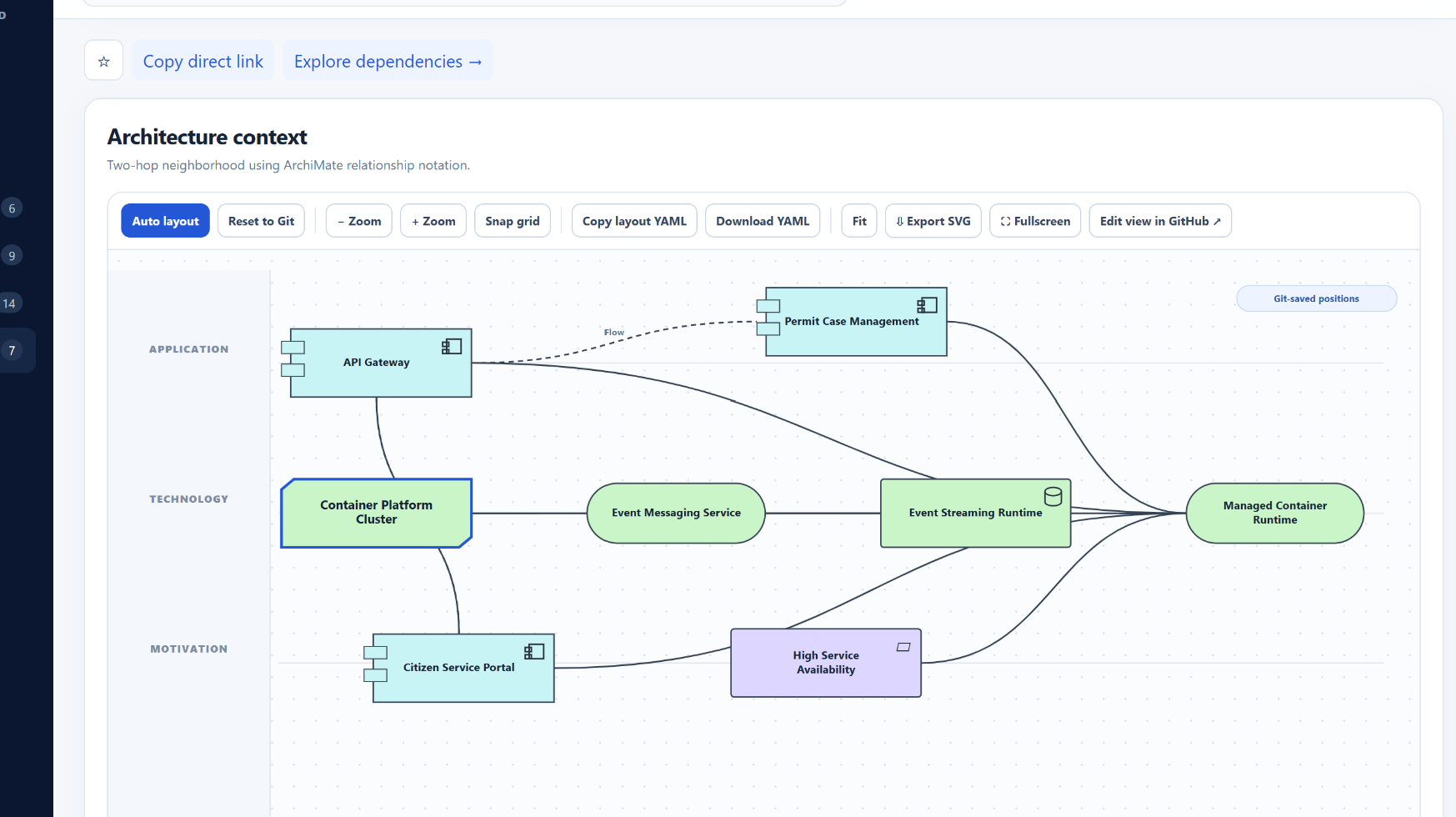}\par\vspace{0.6mm}
        {\footnotesize (d) Interactive architecture context with dependency exploration and layout-as-code controls.}
    \end{minipage}
    \caption{Representative EA-Ops portal views generated from the Metroville reference architecture: (a) repository overview, (b) deterministic governance status, (c) ArchiMate motivation concepts, and (d) cross-layer dependency exploration. All views are derived from the same reviewed YAML objects, relationships, views, and rules used by CI; the screenshots illustrate the implementation and are not used as quantitative evaluation evidence.}
    \label{fig:portal-screens}
\end{figure*}

Figure~\ref{fig:portal-screens} shows representative views generated from the Metroville reference repository. The overview exposes repository-scale facts and quality indicators; the governance page surfaces the same deterministic checks executed in CI and pull requests; the motivation catalog presents architecture intent using ArchiMate motivation concepts; and the architecture-context view provides interactive dependency exploration with Git-backed layout controls. These screens are presentation projections of the reviewed repository model rather than separately maintained architecture data, reducing the risk of divergence between machine-readable facts and stakeholder-facing documentation.

\section{Experimental Methodology}
\label{sec:method}
The portal views in Fig.~\ref{fig:portal-screens} document the implemented user-facing projections of EA-Ops. The empirical claims below are evaluated independently from those screenshots using machine-readable benchmark outputs, ground-truth manifests, raw timing CSVs, and controlled-change oracle comparisons.
\subsection{Research Artifact and Execution Environment}
The evaluation was conducted using EA-Ops version 0.2.0 and the Metroville reference case, both pinned to their latest Git commits. All benchmark jobs were executed via the canonical GitHub Actions research workflow on GitHub-hosted x86-64 runners operating on Ubuntu 24 and Python 3.12.14 with approximately 16 GB of RAM. Hardware allocation varied across scale tiers: the 100-object benchmark ran on an Intel Xeon 6973P-C runner, whereas the 1,000-, 10,000-, and 50,000-object benchmarks executed on AMD EPYC 9V74 runners. Rather than masking this hardware variation through normalization, we explicitly report it as a potential threat to internal timing validity.

\subsection{Synthetic Scalability Models}
The benchmark generator creates deterministic repositories of 100, 1,000, 10,000, 50,000, and 100,000 architecture objects. Each model contains fixed governance fixtures and otherwise consists primarily of ApplicationComponent objects. The relationship count is configured to approximately twice the object count: 200, 2,000, 20,000, 100,000, and 200,000 relationships respectively. Application-to-application Association edges are selected deterministically from a seeded random generator and are semantically valid under the configured profile.

For each completed scale, the workflow executes three warm-up runs followed by 30 measured repetitions. Each measured validation run includes repository loading and full validation. Impact analysis is timed separately using \texttt{app.000000} as the changed node. Report and portal generation are measured for the first three repetitions at each scale because they perform filesystem output and, for the report, increasingly expensive aggregate queries. Raw per-repetition CSV files are retained by the workflow.

For a measured vector $x_1,\ldots,x_n$, the workflow reports the mean, median, standard deviation, 95th percentile, and a normal-approximation 95\% confidence interval for the mean,
\begin{equation}
	\bar{x} \pm 1.96\frac{s}{\sqrt{n}}.
\end{equation}
We emphasize medians and p95 values in the results because hosted cloud runners can exhibit transient scheduling noise.

\subsection{Independent Fault Injection}
RQ1 uses eight fault classes: dangling source, dangling target, invalid ArchiMate relationship, missing owner, missing criticality, duplicate ID, invalid lifecycle value, and organization-governance violation. Each class is executed over 30 deterministic trials, yielding 240 trials in total. Every trial starts from a generated clean 100-object model.

The fault injector does not import the EA-Ops validator. Instead, it performs direct YAML mutations and writes an independent \texttt{ground\_truth.json} manifest containing expected error tuples $(\mathrm{code},\mathrm{object\_id})$. The evaluator compares the complete set of emitted error tuples with ground truth. We calculate
\begin{align}
	\mathrm{Precision} &= \frac{TP}{TP+FP},\\
	\mathrm{Recall} &= \frac{TP}{TP+FN},\\
	F_1 &= 2\frac{\mathrm{Precision}\cdot\mathrm{Recall}}{\mathrm{Precision}+\mathrm{Recall}}.
\end{align}
This experiment measures deterministic detector correctness for the explicitly injected fault catalog; it is not an estimate of prevalence or accuracy on arbitrary real-world EA repositories.

\subsection{Metroville Controlled-Change Case}
The companion repository models the fictional Metroville Digital Permit Service. Its baseline contains 65 objects, 85 relationships, four views, and eight governance rules. The normal CI validation reports a quality score of 100/100, 100\% ownership coverage, and no validation findings.

Ten controlled architecture-change scenarios cover additive changes, lifecycle and ownership changes, removal of a referenced service, classification changes, dangling references, an invalid typed relationship, a new process, and a criticality change. Each scenario is applied to a fresh copy of the base architecture. Validation output is checked against the scenario manifest. Separately, the scenario runner implements an independent breadth-first search over the modified relationship files and compares its complete reachable set with \texttt{eaops.core.impact}. The independent oracle does not call the EA-Ops impact implementation.

\begin{figure*}[t]
	\centering
	\includegraphics[width=0.93\textwidth]{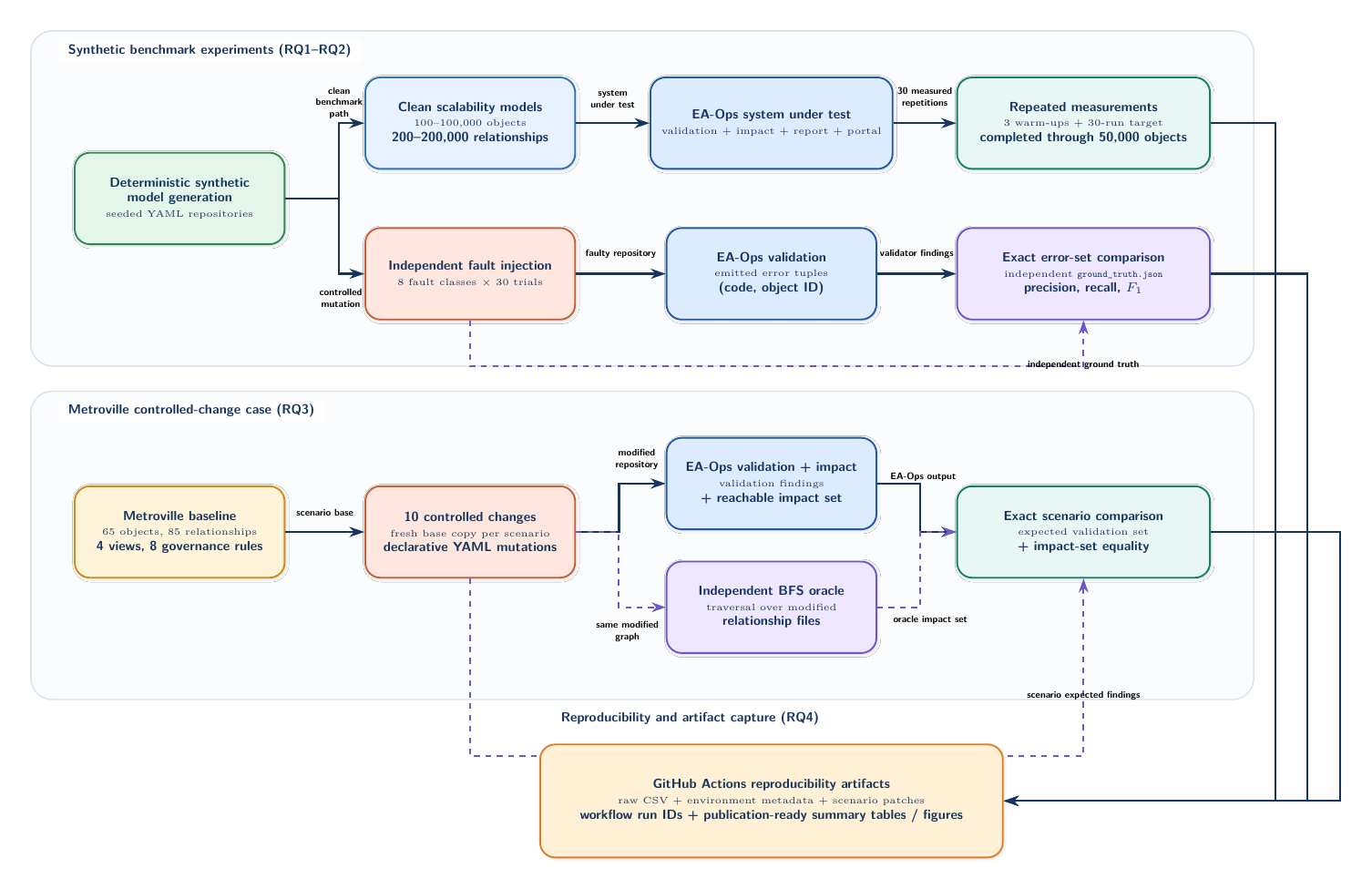}
	\caption{Reproducible evaluation pipeline. Fault ground truth and the Metroville graph oracle are deliberately implemented outside the system-under-test logic.}
	\label{fig:evaluation}
\end{figure*}

\section{Results}
\label{sec:results}
\subsection{RQ1: Validation Correctness}
Table~\ref{tab:faults} reports the fault-injection results. All 30 trials in every fault class produced the exact expected error set. Aggregated over 240 trials, the experiment produced 240 true positives, zero false positives, and zero false negatives, yielding precision, recall, and $F_1$ of 1.000 for the controlled fault catalog.

\begin{table*}[t]
	\centering
	\small
	\caption{Controlled Fault-Injection Accuracy (30 Trials per Fault Class)}
	\label{tab:faults}
	\begin{tabular}{l r r r r r r}
		\toprule
		Fault class & Trials & TP & FP & FN & Precision & Recall / $F_1$ \\
		\midrule
		Dangling source                & 30 & 30 & 0 & 0 & 1.000 & 1.000 / 1.000 \\
		Dangling target                & 30 & 30 & 0 & 0 & 1.000 & 1.000 / 1.000 \\
		Invalid ArchiMate relationship & 30 & 30 & 0 & 0 & 1.000 & 1.000 / 1.000 \\
		Missing owner                  & 30 & 30 & 0 & 0 & 1.000 & 1.000 / 1.000 \\
		Missing criticality            & 30 & 30 & 0 & 0 & 1.000 & 1.000 / 1.000 \\
		Duplicate identifier           & 30 & 30 & 0 & 0 & 1.000 & 1.000 / 1.000 \\
		Invalid lifecycle              & 30 & 30 & 0 & 0 & 1.000 & 1.000 / 1.000 \\
		Governance violation           & 30 & 30 & 0 & 0 & 1.000 & 1.000 / 1.000 \\
		\midrule
		\textbf{Total} & \textbf{240} & \textbf{240} & \textbf{0} & \textbf{0} & \textbf{1.000} & \textbf{1.000 / 1.000} \\
		\bottomrule
	\end{tabular}
\end{table*}

The result is important primarily as a reproducibility and regression result: the validator's structural, semantic, and configured governance findings are deterministic and agree with independently specified expectations for these fault types. It should not be interpreted as evidence that eight fault classes cover all possible EA-quality defects.

\subsection{RQ2: Scalability}
\begin{figure}[t]
	\centering
	\includegraphics[width=\columnwidth]{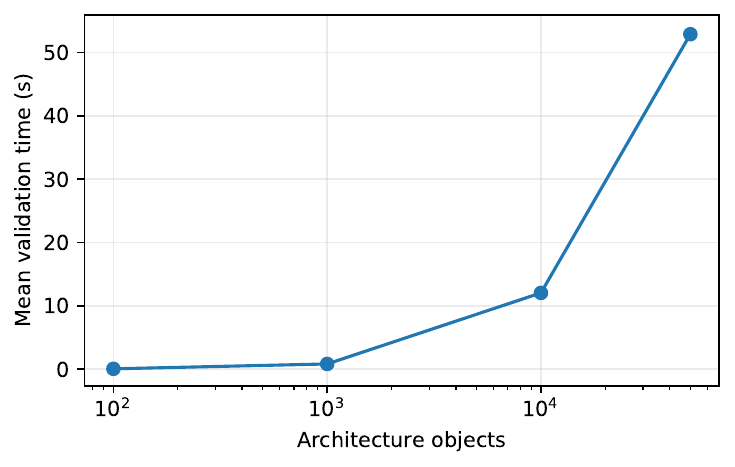}
	\caption{Mean end-to-end validation time with normal-approximation 95\% confidence intervals over 30 measured repetitions.}
	\label{fig:valscale}
\end{figure}

Table~\ref{tab:scale} summarizes the completed performance experiments. Validation median time increased from 0.071~s for 100 objects and 200 relationships to 52.582~s for 50,000 objects and 100,000 relationships. Peak resident memory increased from 31.9~MB to 919.2~MB. Mean validation times and 95\% confidence intervals are visualized in Fig.~\ref{fig:valscale}.

\begin{table*}[t]
	\centering
	\small
	\caption{EA-Ops Scalability Results. Validation and Impact Use 30 Measured Repetitions; Report and Portal Use 3 Measured Repetitions.}
	\label{tab:scale}
	\setlength{\tabcolsep}{5pt}
	\resizebox{\textwidth}{!}{%
\begin{tabular}{r r r r r r r r}
		\toprule
		\textbf{Objects} & \textbf{Relationships} & 
		\begin{tabular}[c]{@{}c@{}}\textbf{Validation}\\\textbf{median (s)}\end{tabular} & 
		\begin{tabular}[c]{@{}c@{}}\textbf{Validation}\\\textbf{p95 (s)}\end{tabular} & 
		\begin{tabular}[c]{@{}c@{}}\textbf{Impact}\\\textbf{median (ms)}\end{tabular} & 
		\begin{tabular}[c]{@{}c@{}}\textbf{Peak RSS}\\\textbf{(MB)}\end{tabular} & 
		\begin{tabular}[c]{@{}c@{}}\textbf{Report}\\\textbf{median (s)}\end{tabular} & 
		\begin{tabular}[c]{@{}c@{}}\textbf{Portal}\\\textbf{median (s)}\end{tabular} \\
		\midrule
		100    & 200     & 0.071  & 0.076  & 0.118   & 31.9  & 0.001   & 0.001 \\
		1,000  & 2,000   & 0.842  & 0.858  & 2.691   & 48.2  & 0.085   & 0.016 \\
		10,000 & 20,000  & 12.034 & 12.506 & 89.814  & 220.4 & 19.157  & 0.218 \\
		50,000 & 100,000 & 52.582 & 55.047 & 627.000 & 919.2 & 823.531 & 0.898 \\
		\bottomrule
	\end{tabular}%
}
\end{table*}

\begin{figure}[t]
	\centering
	\includegraphics[width=\columnwidth]{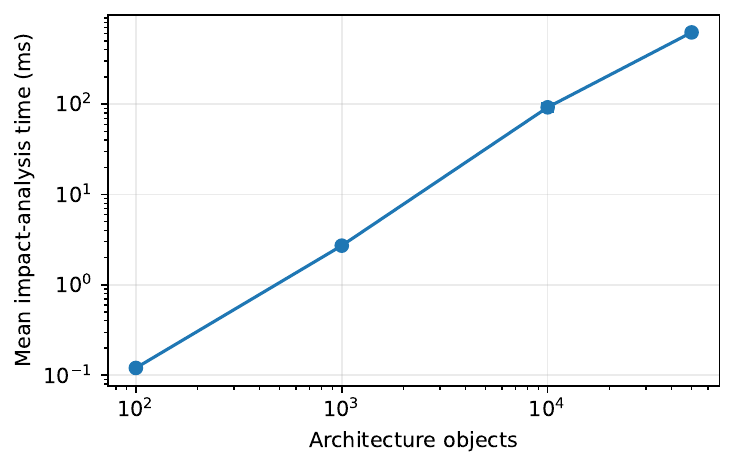}
	\caption{Mean change-impact latency with 95\% confidence intervals. The y-axis is logarithmic.}
	\label{fig:impactscale}
\end{figure}

Impact analysis remained below one second at the largest completed scale: the median increased from 0.118~ms at 100 objects to 627.000~ms at 50,000 objects (Fig.~\ref{fig:impactscale}). The synthetic graphs form large connected components: the median impacted set was 94 of 100 nodes, 979 of 1,000, 9,781 of 10,000, and 49,045 of 50,000. Thus the impact experiment is not a trivial local-neighborhood lookup; it traverses almost the entire connected model.

\begin{figure}[t]
	\centering
	\includegraphics[width=\columnwidth]{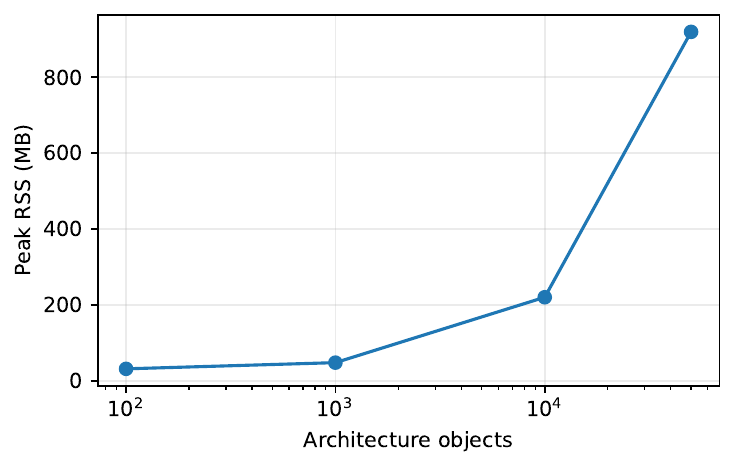}
	\caption{Maximum observed process peak RSS for each completed scalability job.}
	\label{fig:memory}
\end{figure}

Figure~\ref{fig:memory} shows memory growth. The 50,000-object experiment remained well within the approximately 16~GB runner capacity, with a measured peak RSS below 1~GB.

\begin{figure}[t]
	\centering
	\includegraphics[width=\columnwidth]{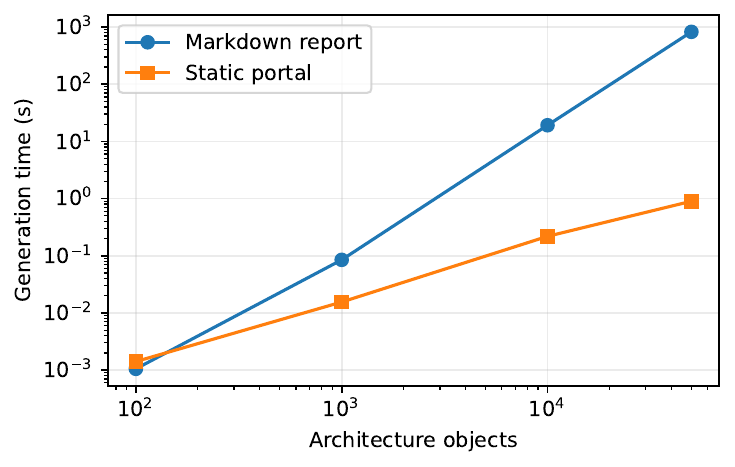}
	\caption{Median publication-generation time. Both axes use logarithmic scaling where applicable; report generation becomes the dominant bottleneck.}
	\label{fig:publication}
\end{figure}

The main scalability bottleneck is not semantic validation or portal serialization but Markdown report generation. At 50,000 objects, the median report time was 823.5~s (13.7~min), whereas static portal generation took 0.898~s after the model was loaded (Fig.~\ref{fig:publication}). Inspection of the current report implementation explains this behavior: application rows repeatedly scan the relationship collection to count support relationships, yielding $O(|A||E|)$ work for an application-heavy synthetic model. This is a concrete optimization target rather than a limitation of the architecture-as-code representation itself.

The workflow also generated a 100,000-object / 200,000-relationship repository and captured its environment metadata. However, the performance job did not complete within the configured 180-minute job budget while executing the benchmark, and therefore no valid 100,000-object timing row was emitted. We treat 50,000 objects as the largest fully measured scale and the 100,000-object timeout as an observed end-to-end boundary. No value is extrapolated or imputed for the incomplete run.

\subsection{RQ3: Metroville Controlled Changes}
All ten Metroville scenarios matched both their validation expectations and the independent breadth-first-search impact oracle. Table~\ref{tab:metro} summarizes the scenarios. Eight scenarios were expected to remain semantically valid; CR04 deliberately removed a referenced identity service and generated the two expected missing-endpoint findings, while CR07 and CR08 generated the expected dangling-target and invalid-relationship findings respectively. In every case the actual validation set matched exactly.

\begin{table*}[t]
	\centering
	\small
	\caption{Metroville Controlled Architecture-Change Results}
	\label{tab:metro}
	\resizebox{\textwidth}{!}{%
\begin{tabular}{l p{5.5cm} p{4.8cm} r c}
		\toprule
		\textbf{ID} & \textbf{Controlled change} & \textbf{Expected validation finding(s)} & \textbf{Impact size} & \textbf{Exact oracle match} \\
		\midrule
		CR01 & Add dedicated payment application service & None & 49 & Yes \\
		CR02 & Mark notification application for elimination & None & 48 & Yes \\
		CR03 & Change permit-assessment process owner & None & 48 & Yes \\
		CR04 & Remove identity service but retain references & MISSING\_SOURCE; MISSING\_TARGET & 48 & Yes \\
		CR05 & Move container platform to migration lifecycle & None & 48 & Yes \\
		CR06 & Raise citizen-profile classification & None & 48 & Yes \\
		CR07 & Introduce dangling target in payment flow & MISSING\_TARGET & 49 & Yes \\
		CR08 & Replace application flow with invalid Access edge & INVALID\_RELATIONSHIP & 48 & Yes \\
		CR09 & Add governed appeal-review process & None & 49 & Yes \\
		CR10 & Promote payment process criticality & None & 48 & Yes \\
		\bottomrule
	\end{tabular}%
}
\end{table*}

The impact-set sizes of 48 or 49 reflect the connectivity of the modeled service rather than an imposed threshold. The independent oracle and EA-Ops agreed on every reachable ID. This provides stronger evidence than simply demonstrating that the command executes: the test checks set equality against an implementation outside the EA-Ops impact code.

\subsection{RQ4: Reproducibility}
The evaluation is implemented in the same Git-native medium advocated by the approach. The research workflow defines the scale matrix, seeds, warm-ups, repetitions, fault catalog, artifact retention, and aggregation steps. Every performance job records the Git commit, workflow run ID, OS image, Python version, CPU model, memory, and runner metadata. Raw CSV files are uploaded as workflow artifacts. The Metroville workflow additionally emits a unified patch for each controlled scenario, making the exact model mutation auditable.

This does not eliminate all environmental variation, but it makes the variation visible and the experiment re-executable. The research harness therefore serves two roles: an empirical evaluation of EA-Ops and an example of how architecture governance itself can be subjected to reproducible CI.

\section{Discussion}
\label{sec:discussion}
\subsection{What the Results Support}
The fault experiment supports a narrow but strong claim: for the tested structural, semantic, and policy faults, EA-Ops behaved deterministically and exactly matched independently declared expectations across repeated generated models. The Metroville experiment supports a complementary claim at the repository level: controlled cross-layer changes can be applied as code, evaluated by the same architecture gate used in CI, and checked against an independent graph oracle.

The scalability results show that semantic validation and broad connected-component impact analysis remain operationally practical at tens of thousands of objects on commodity hosted CI runners. A median validation time of 52.6~s for 50,000 objects and 100,000 relationships is compatible with a pull-request quality gate in many organizations, although acceptable latency is context dependent. The much slower Markdown report generation reveals that peripheral presentation algorithms can dominate end-to-end scale even when the underlying validator remains usable.

These findings reinforce a broader interpretation of Architecture as Code: the value is not merely textual serialization. The useful engineering property comes from coupling machine-readable architecture semantics with repeatable validation, executable organizational policy, versioned change review, and derivation of human-facing views from the same source.

\subsection{Practical Implications}
EA-Ops deliberately reuses standard Git hosting rather than implementing users, roles, approvals, and audit history inside a new EA database. This can reduce integration surface for organizations that already operate Git-based engineering platforms. Architecture teams can adopt progressive governance: begin with stable IDs and required ownership, add semantic relationship checks, then introduce policy rules and automated impact summaries as the repository matures.

The evaluation also suggests a practical separation of concerns. Semantic validation should remain lightweight and PR-oriented; expensive reports should be indexed or pre-aggregated rather than recomputed with repeated full-edge scans. Similarly, impact analysis should evolve from the current undirected connected set toward purpose-specific semantics, such as directional dependency impact, relationship-type filters, bounded hop distance, or risk-weighted propagation.

\subsection{Threats to Validity}
\textbf{Construct validity:} The fault catalog measures eight explicit classes and exact emitted tuples. It does not measure whether the rules capture all important architecture-quality concerns. Likewise, graph reachability is a structural impact proxy, not proof of business or operational consequence.

\textbf{Internal validity:} Fault ground truth and the Metroville impact oracle are implemented independently of the corresponding EA-Ops validator and impact functions, reducing circularity. However, both run in the same Python/CI environment and can share assumptions about YAML structure. The synthetic generator intentionally emits models compatible with the configured metamodel; malformed serialization beyond the selected mutations is not studied.

\textbf{Performance validity:} GitHub-hosted runners are not hardware-pinned. One completed scale ran on an Intel CPU and the larger scales on AMD EPYC CPUs. Repeated measurements, warm-ups, raw data, and confidence intervals reduce but do not remove this threat. Peak RSS is process-level operating-system telemetry and should be interpreted as an implementation measurement rather than a precise allocation profile.

\textbf{External validity:} The synthetic topology is application-heavy and contains approximately two relationships per object. Real EA repositories may have different layer distributions, relationship density, modularity, and policy complexity. Metroville is fictional and intentionally compact; it does not constitute evidence from a production city administration. A future multi-organization study should examine review effort, governance adoption, and model evolution over time.

\textbf{Semantic validity:} The built-in profile checks relationship combinations against a pinned ArchiMate 3.2 matrix, but EA-Ops is not claimed to be an Open Group certified tool and does not implement every possible normative constraint. The current impact algorithm also treats relationships as undirected for reachability. These boundaries are stated explicitly to avoid conflating useful semantic checking with complete language conformance.

\section{Open Science and Reproducibility}
\label{sec:repro}
The EA-Ops framework is publicly available at
\url{https://github.com/vtavakkoli/ea-ops} under the Apache-2.0
license. The Metroville reference architecture and the controlled-change
scenarios are available at
\url{https://github.com/vtavakkoli/ea-ops-example}.

The benchmark scripts generate the evaluation models deterministically from
documented seeds and do not require large generated datasets to be stored in
the repository. The accompanying repositories contain the source code,
benchmark configurations, controlled scenarios, and CI workflow definitions
needed to reproduce the experiments described in this article.

Because the complete experimental procedure is maintained as executable,
version-controlled workflow definitions, reproduction does not depend on a
particular GitHub Actions run identifier or individual commit number. The
repositories provide the implementation and evaluation infrastructure required
to repeat the experiments and inspect the generated results. For long-term
archival reproducibility, a tagged research release may additionally be
deposited in a persistent software archive.

\section{Conclusion}
\label{sec:conclusion}
This work presented EA-Ops, a Git-native Enterprise Architecture-as-Code framework that integrates YAML-based architecture facts, ArchiMate-aware relationship validation, organization-specific governance rules, pull-request review, graph-based impact analysis, and static publication. The evaluation produced three main empirical observations. First, eight independently injected fault classes were detected with exact precision, recall, and $F_1$ of 1.000 across 240 controlled trials. Second, fully measured scalability experiments reached 50,000 objects and 100,000 relationships, where median validation was 52.6~s, median impact traversal was 627~ms, and peak RSS remained below 1~GB. Third, all ten controlled Metroville change scenarios exactly matched both expected validation outcomes and an independent impact oracle.

The evaluation also exposed a concrete limit: the 100,000-object end-to-end job exceeded the 180-minute CI budget, with report generation already dominating runtime at 50,000 objects. This negative result points directly to the next engineering step: indexed report aggregation and adjacency-indexed or semantically typed impact traversal. More broadly, the findings indicate that treating enterprise architecture as a reviewed, executable, and reproducible engineering artifact is feasible without abandoning established EA semantics. Future work should validate the approach in production organizations, compare review effort with conventional EA tooling, and study typed impact propagation and longitudinal model evolution.

\paragraph{Vahid Tavakkoli.}
received the M.Sc. degree from Carinthia University of Applied Sciences, Austria, and the Dr. techn. degree from the University of Klagenfurt, Austria. He is affiliated with the Department of Smart Systems Technologies, University of Klagenfurt. His research interests include artificial intelligence, machine learning, distributed and multi-agent systems, autonomous systems, and reproducible computational methods.

\paragraph{Kabeh Mohsenzadegan.}
received the Dr. techn. degree in information technology from the University of Klagenfurt, Klagenfurt, Austria, in 2023. Mohsenzadegan is affiliated with the Department of Smart Systems Technologies, University of Klagenfurt. Research activities include artificial intelligence, multi-agent and autonomous systems, knowledge-based systems, and intelligent transportation and smart-system applications.

\paragraph{Kyandoghere Kyamakya.}
received the Ingenieur Civil degree in electrical engineering from the University of Kinshasa, Democratic Republic of the Congo, in 1990, and the Ph.D. degree in electrical engineering from the University of Hagen, Germany, in 1999. After postdoctoral research in mobility management for wireless networks at Leibniz University Hannover, Germany, he served there as a Junior Professor of positioning and location-based services from 2002 to 2005. Since 2005, he has been a Full Professor with the University of Klagenfurt, Austria, where he heads research in transportation informatics. His research interests include machine learning, modeling and simulation, intelligent transportation systems, robotics, telecommunications, navigation, and intelligent logistics.

\end{document}